\documentclass{appolb}
\usepackage{epsfig}
\usepackage{amsmath}
\usepackage{enumitem}

\begin{document}
\title{Solving the puzzle of nuclear stopping power
}
\author{Marek Je\.zabek$^1$ and Andrzej Rybicki$^1$
\address{$^1$Institute of Nuclear Physics, Polish 
 Academy of Sciences, \\
Radzikowskiego 152, 31-342~Krak\'ow, Poland}
}
\maketitle
\begin{abstract}
This paper reviews our recent findings on the dynamics of transport of baryon number in proton-induced reactions at the CERN SPS. These are put in a more general context of the present understanding of baryon stopping phenomena up to LHC and cosmic ray energies. The implications of our studies, advantages to be provided by modern and high quality experimental data, and perspectives of new measurements are shortly discussed.
 \end{abstract}



\section{Introduction}
\label{sec1}

Obtaining quantitative predictions for numerous phenomena and processes in particle and nuclear physics as well as astrophysics is not possible without a reliable phenomenological description of momentum distributions of baryons emitted in ``soft'' multiparticle processes. Achieving such a description is therefore an important problem in high energy physics. This problem emerged several decades ago, at the very beginning of studies of collisions of high energy cosmic rays with atomic nuclei in Earth's atmosphere. For the inelastic nucleon-nucleus interactions, by soft we mean processes where transverse momenta of produced particles are of the order of a few hundreds of MeV. At beam energies of hundreds of GeV such processes dominate the inelastic cross section. Unfortunately, perturbative chromodynamics (pQCD) does not provide reliable predictions for soft processes. For that reason the construction of a reliable phenomenological model, describing the spectra of emitted baryons at beam energies from a few tens of GeV upwards constitutes, still at the present time, a serious challenge which calls for a unified experimental and theoretical effort.

It should be stated that for many decades, the available experimental data were not particularly restrictive. Consequently they could be described by models based on very different, and often contradictory assumptions. For proton-proton ($pp$) collisions, quite a reasonable starting point was the flat Feynman-x ($x_F$) distribution of protons in the final state. This implied that after colliding with another nucleon the proton lost, on the average, half of its initial energy. However, in processes of cosmic ray proton scattering on atmospheric nitrogen or oxygen nuclei, the proton may often interact with two or more nucleons in the nucleus. Very popular still in the late 1970's was a {\em sequential description}, in which the proton energy was degrading so that every next proton-nucleon collision was occurring at a correspondingly lower projectile energy. It should be underlined that this simple description worked pretty well for the experimental data available at the time. This is notwithstanding that it is contradicted by other well documented facts, in particular by the $A$-dependence of ``hard'' processes with large transfer of four-momentum, {\em e.g.} for lepton-pair production in the Drell-Yan process (see, {\em e.g.}, Ref.~\cite{jj}). Indeed, if the projectile would loose a significant part of its initial energy in between the collisions with subsequent nucleons, the increase of corresponding cross-sections with $A$ would be much slower than observed experimentally. The same conclusion can be obtained from the comparison of time intervals between the collisions with subsequent ``wounded nucleons'' and the ``formation time'' required for production of fast secondary particles in soft processes. For high enough energies of the projectile, the size of the corresponding ``formation zone'' of such fast particles exceeds the (longitudinal) size of the Lorentz-contracted nucleus~\cite{m-blan,m-b2,m-b1}. 
%
%
In Ref.~\cite{bialas-stodolsky-apb} this idea was connected to the ``bremsstrahlung analogy'' by Stodolsky~\cite{stodolsky}. The latter linked together the increase of multiplicity of produced particles with $\ln{s}$, the presence of plateau in their rapidity, and the energy loss spectrum of the incident particle. For $pA$ collisions~\cite{bialas-stodolsky-apb}, the bremsstrahlung analogy predicted the plateau to be ``somewhat higher'' than in $pp$ collisions, and an increase in leading particle inelasticity. It also predicted a {\em qualitative} although not {\em quantitative} similarity of leading particle spectra in $pp$ and $pA$ collisions.

\begin{figure}[t]
\vspace{0.2cm}
\begin{center}
\hspace*{-1.3cm}
\includegraphics[width=10.2cm]{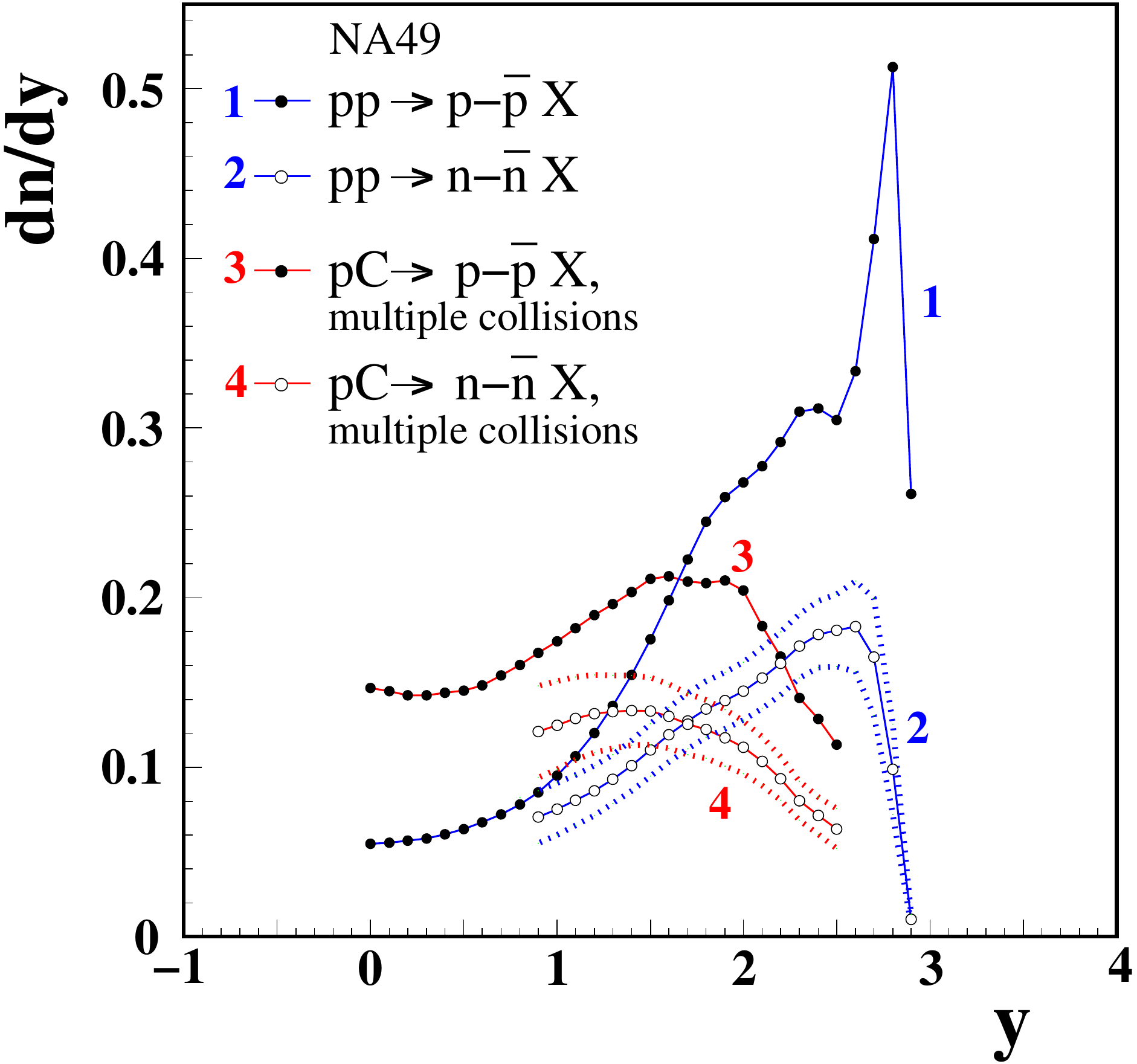}\\
\end{center}
  \caption{Rapidity distributions of net protons ($p-\overline{p}$) and net neutrons ($n-\overline{n}$) in $pp$ collisions and in $pC$ reactions in which the projectile proton collides with more than one target nucleon. Both reactions are taken at $\sqrt{s}_\mathrm{_{NN}}=17.3$~GeV. Both sets of distributions have been obtained from NA49 data~\cite{x,y}, see Ref.~\cite{g} for a detailed description. The distribution no. 1 was first presented in Ref.~\cite{x} and the other distributions in Ref.~\cite{g}. 
The average number of proton-nucleon collisions in this ``multiple collision'' $pC$ sample is about 2.6~\cite{pcdiscus}.
The dotted curves reflect the neutron systematic error which is the main source of systematic uncertainty in the data~\cite{x,y}.}
 \label{fig1}
\vspace{0.2cm}
\end{figure}

It is interesting to revisit this 
idea with modern experimental data~\cite{x,y}, shown in Fig.~\ref{fig1}. 
It seems evident to us that the postulated qualitative similarity is seriously challenged by the experiment. The simplest example of a {\em qualitative difference} is the ``diffractive'' proton peak at high 
rapidity
in $pp$ collisions, which is completely gone in $pC$ reactions where
 the projectile proton collides with more than one nucleon. The interested Reader is also invited to inspect the corresponding proton $x_F$ spectrum in Fig.~2 in our Ref.~\cite{g}, 
which is
in sharp contrast 
with the predicted distributions in Fig.~3 of Ref.~\cite{bialas-stodolsky-apb}. The reasons which we expect for 
the presence of
such {\em qualitative differences} will be addressed in Secs.~\ref{sec3} and~\ref{sec35}.

It is evident in Fig.~\ref{fig1} that the projectile perfectly ``remembers'' the number of wounded nucleons that it crosses during the collision, and that it ``responds'' to it by a large increase in nuclear stopping power\footnote{Term taken from Ref.~\cite{Busza} where the increase discussed in the text was seen much earlier.}. A realistic phenomenological model has therefore to explain how the description of the proton {\em encodes} the information on the number of crossed wounded nucleons, as this information cannot be encoded by the total projectile energy as we explained above. A simple
%
%
%
realization of this {\em encoding} idea is the Wounded Quark Model~\cite{bialasczyz}, in which the number of ``wounded'' quarks in the projectile is correlated with the number of wounded nucleons in the target nucleus. The correlation is not unambiguous, as the number of projectile wounded quarks cannot be larger than three for protonic nor larger than two for mesonic beams. 
The same is valid for the
more recent implementation of the same idea which is the Wounded Constituent Model, where the wounded quarks are being replaced by the ``wounded'' 
or 
``unwounded'' quark and diquark~\cite{bialasbzdak1,bialasbzdak2,bialasbzdak3}.

Another possibility is the Dual Parton Model (DPM) by Capella and Tran Thanh Van~\cite{CapellaTranh}, where for the case of $n$ wounded target nucleons the projectile proton is described by a set of $2n$ partons: one diquark, one valence quark and $(n-1)$ pairs of sea quarks and antiquarks. In this approach the description of the projectile is, by definition, unambiguously connected to the number of wounded nucleons in the target. 
As DPM postulates the formation of $2n$ strings (``chains'') between partons from the projectile and these from the wounded nucleons,
the increase of central produced particle density with $n$ is an inherent, natural feature of this model\footnote{We note here the similarity between DPM and the parton model approach by Brodsky, Gunion and K\"uhn~\cite{bgk}, where the ``wee'' partons from the projectile interacted with the ``wee'' partons from essentially independent target nucleons.}. 
It seems quite obvious that both 
DPM and Wounded Quark/Wounded Constituent
models should give similar predictions for light nuclei and not too large values of $n$, and for such characteristics of final state particles for which their $n$-dependence is not very strong. 
In the subsequent discussion we will limit ourselves to the DPM and similar models, because it seems to us that this class of models has a higher potential for the description of strongly $n$-dependent experimental data than the Wounded Quark or Wounded Constituent approaches.
As it was already apparent in Fig.~\ref{fig1}, such a strong $n$-dependence is characteristic of baryon transport phenomena. As we will further discuss below, the enrichment of the phenomenological description by the addition of sea quarks and antiquarks brings new, interesting options which can be confronted with the experiment. 


\section{The puzzle of nuclear stopping power}
\label{puzzle}

It should be clearly underlined that longitudinal momentum conservation, implemented explicitly in the Dual Parton Model, results in softening of diquark and consequently baryon distributions with increasing $n$; this issue will be discussed in detail in Sec.~\ref{sec2} below. {However}, as it was demonstrated by Je\.zabek and R\'o\.za\'nska in Ref.~\cite{comments}, the corresponding effect on the nuclear stopping power was far smaller than what was obtained in the experimental data-based analysis by Busza and Goldhaber~\cite{Busza}. The paradox consisted in the fact that the 
latter result
could be easily explained in the sequential description 
of energy degradation in 
subsequent proton-nucleon collisions in the nucleus, 
which was
contradicted by well documented facts and arguments based on the formation time 
addressed in Sec.~\ref{sec1} above. More realistic models like DPM predicted a much weaker softening of leading particle (baryon) spectra in $pA$ collisions. This paradox we will refer to as ``the puzzle of nuclear stopping power''. As we will demonstrate in this paper which summarizes the implications from our recent works~\cite{g,e,b}, 
now
this puzzle has been solved
with the help of modern experimental data~\cite{x,y}. As we discuss it below, the key elements necessary for solving this puzzle are a rigorous treatment of the role of the color quantum number, and a properly complete description of the proton/nucleon {\em substructure} in terms of available Fock states. Sea partons appear to play an essential role in 
building the nuclear 
 stopping power.
An important practical 
issue appears to be a
proper treatment of {\em isospin effects} in baryon stopping, 
the lack of which we consider as one of the reasons why the
description of this process made below was not formulated 
before
on the basis of
earlier experimental data.

\section{The Dual Parton Model {\em versus} modern experimental data}
\label{sec2}

While this issue was already 
discussed
in our three precedent works~\cite{g,e,b}, one should again 
address
the importance of modern experimental data~\cite{x,y} for the study of baryon transport processes which we review here. The completeness of this data lifts up several of the limitations inherent to earlier studies of this type~\cite{Busza,comments,MJZalewski,Brenner,Bailey,MJ1985,buszaledoux}, 
up to a level which we find incomparable to any other data set known to us. The mere fact that the same experiment provides information on simultaneously proton and neutron spectra is already an inestimable advantage as it frees the study from the extremely risky necessity to use {\em protons} as a proxy for {\em the total baryon number}, which prompts the phenomenologist to model-dependent assumptions that can easily appear highly erroneous with respect to reality. In fact we have been able to verify and falsify some of the earlier assumptions~\cite{MJZalewski}, and found that baryon emission at high rapidities is characterized by strong isospin effects which are very informative on the fate of valence partons in the collision, see Refs.~\cite{g,b} for more details. The general conclusion from our studies is that the earlier data sets were not restrictive enough to make real use of baryon number conservation neither in $pp$ not in $pA$ collisions, an issue which we will further elaborate upon in Sec.~\ref{sec4}.

\begin{figure}[t]
\vspace{0.2cm}
\begin{center}
\hspace*{-1.6cm}
\includegraphics[width=9.2cm]{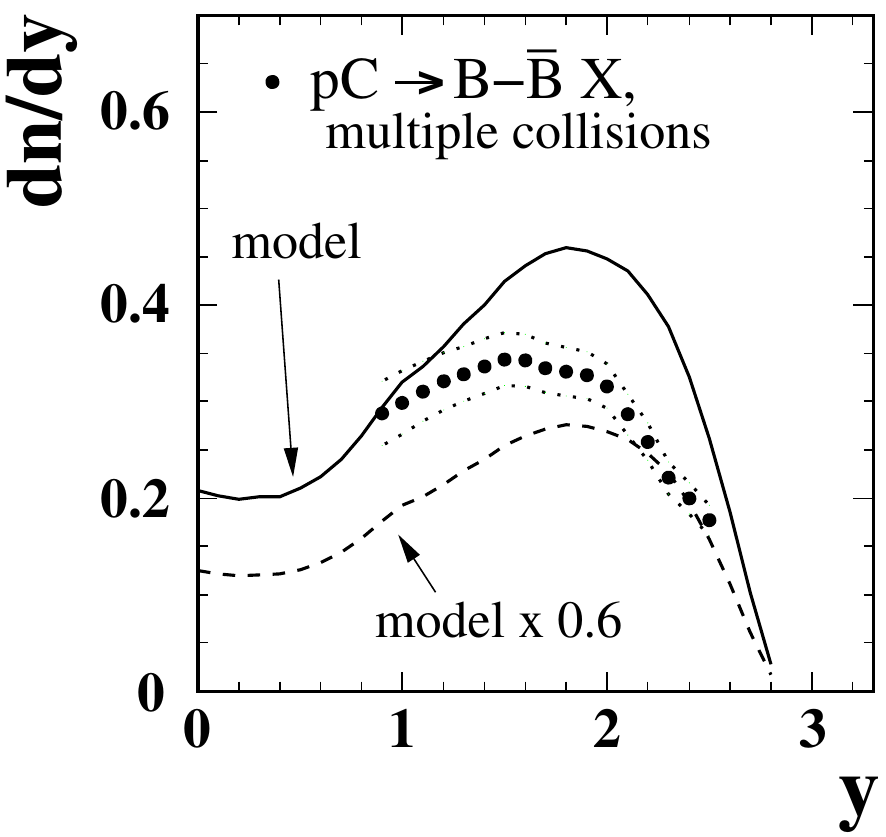}\\
\end{center}
  \caption{Rapidity distribution of net non-strange baryons (summed net protons and net neutrons) in $pC$ reactions in which the projectile proton collides with more than one target nucleon, compared to our DPM simulation described in the text. The figure comes from Ref.~\cite{e}. The dashed curve illustrates the result of our DPM calculation (solid line) scaled by 0.6 which gives the upper limit for the contribution of the diquark-preserving mechanism by Capella and Tran Thanh Van to the total net baryon spectrum.}
 \label{X2}
\vspace{0.2cm}
\end{figure}

\begin{figure}[t]
\vspace{0.2cm}
\begin{center}
\hspace*{-1.1cm}
\includegraphics[width=8.2cm]{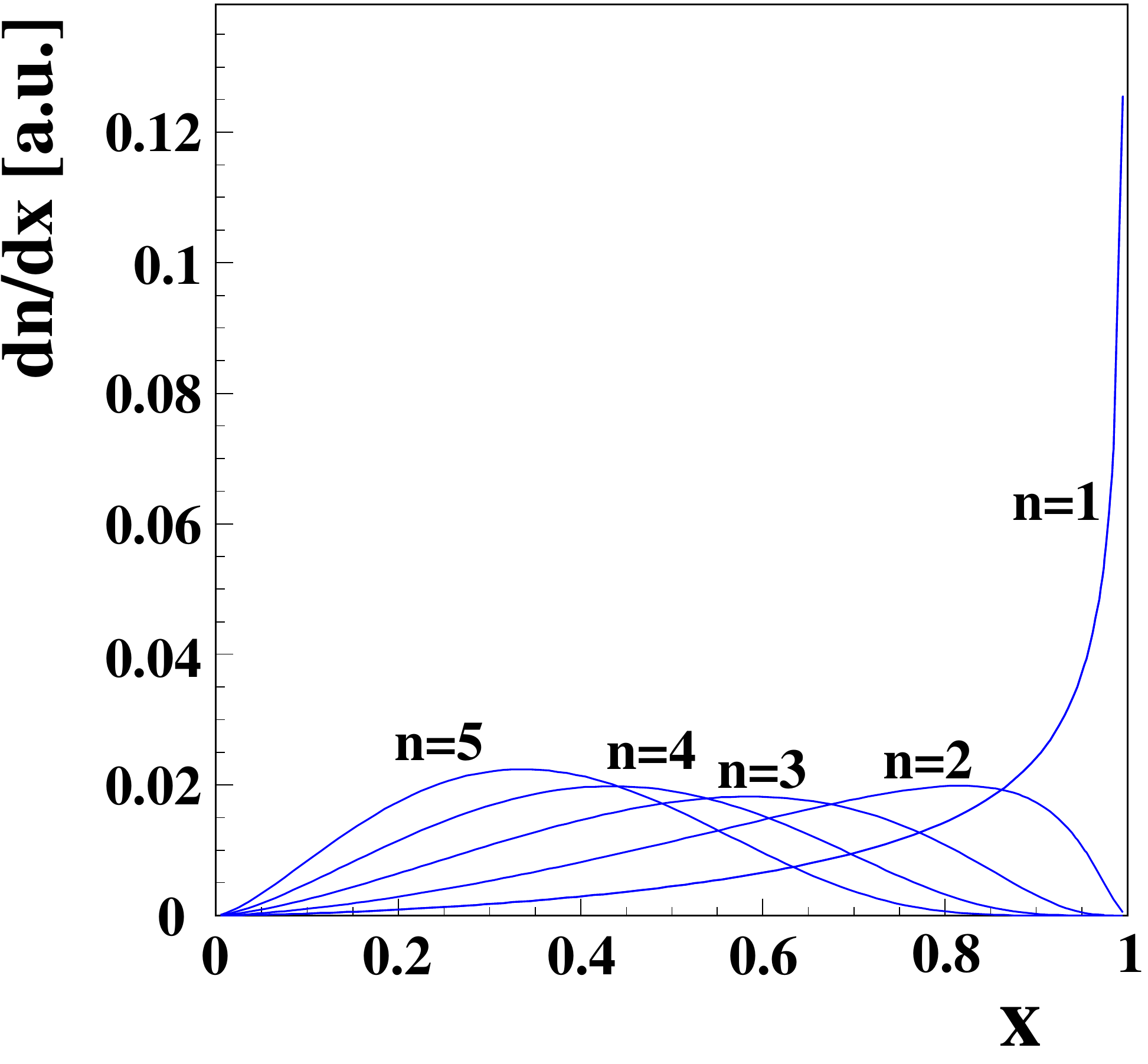}\\
\end{center}
\vspace*{-0.1cm}
  \caption{Momentum fraction $x$ distribution for the projectile diquark in $pA$ collisions for $n=1, 2, ...$, obtained assuming the original mechanism by Capella and Tran Thanh Van as described in the text. The ``multiple collision'' $pC$ sample shown in Figs.~\ref{fig1} and~\ref{X2} correspond mostly to $n=2$ and $n=3$.}
 \label{X3}
\vspace{0.2cm}
\end{figure}

In our analysis of the above-mentioned precise NA49 data we demonstrated~\cite{e}, that the DPM is not able to provide a correct description for the distribution of net baryon number for ``multiple'' proton-nucleon collisions\footnote{That is, $pC$ reactions where the projectile proton collides with more than one nucleon from the target, see Ref.~\cite{e} for more details.}. This is shown in Fig.~\ref{X2}.
The failure of the model takes place {\em in spite} of the fact that the corresponding baryon number distribution in $pp$ collisions is very well tuned to 
experimental $pp$ data. 

With the diquark and quark momentum distributions postulated by the DPM~\cite{CapellaTranh,MJ1985}, 
the above 
tuning consists in defining the proper fragmentation function of the diquark-quark string (``chain'') on the basis of experimental distributions of 
baryons
in $pp$ collisions; 
details can be found in Ref.~\cite{b}. Consequently, for multiple proton-nucleon collisions in inelastic proton scattering on the atomic nucleus, for $n$ wounded nucleons which results in the addition of $(n-1)$ sea quark-antiquark pairs to the set of projectile constituents, the problem reduces to the following: the original mechanism introduced by Capella and Tran Thanh Van results in the decrease of the diquark's longitudinal momentum, {\em but to a degree far smaller than what is required in order to obtain a proper description of the data} (Fig.~\ref{X2}). In Fig.~\ref{X3} we present the longitudinal momentum distributions of the diquark for a few values of $n$. 
As it appears the changes with increasing number of wounded nucleons are not
very 
large, in particular for $n=2$ and $n=3$ which built up most of the ``multiple collision'' $pC$ sample shown in Fig.~\ref{X2}. 
This implies, as it was already noticed in Ref.~\cite{comments}, that the DPM predicts a much weaker stopping of secondary baryons (the nuclear stopping power)
than the naive sequential model with losses of energy in between the collisions, {\em and also} the nuclear stopping power obtained from the analysis of FNAL experimental data~\cite{Busza}. Quite paradoxically, this can be considered as ``good'' news for the DPM as it does not describe the experimental data but at the same time it is not self-contradictory as the naive sequential model. The distributions in Fig.~\ref{X3} suggest as well that the predictions from the DPM should not be very different from predictions of the Wounded Quark or Wounded Constituent Models. Consequently, we find it possible that the latter will have a similar problem with the description of baryon stopping in multiple proton-nucleon collisions.

\section{Strong baryon stopping}
\label{sec3}

The discussion made above leads to the following question: how to explain the effect of the strong softening of the projectile (the strong nuclear stopping power) in multiple proton-nucleon collisions, once the momentum distributions of constituent partons cannot change during the (short) time which the latter spends on passing through the nucleus?

The answer which we propose is the following: in the projectile hemisphere (in the proton-nucleon c.m.s.), the secondary baryon is created from the constituents of the projectile and of 
nucleons from
the nucleus. What kind of system that is and which constituents belong to it depends on the collision 
we are dealing with. For instance, for the baryon originating from the fragmentation of the diquark-quark chain, Fig.~\ref{X2}, the {\em momenta} and {\em isospins (flavors)} of the ``ends'' of this chain will depend, among others, on the number $n$ of wounded nucleons. 

\begin{figure}[t]
\vspace{0.2cm}
\begin{center}
\includegraphics[width=10.2cm]{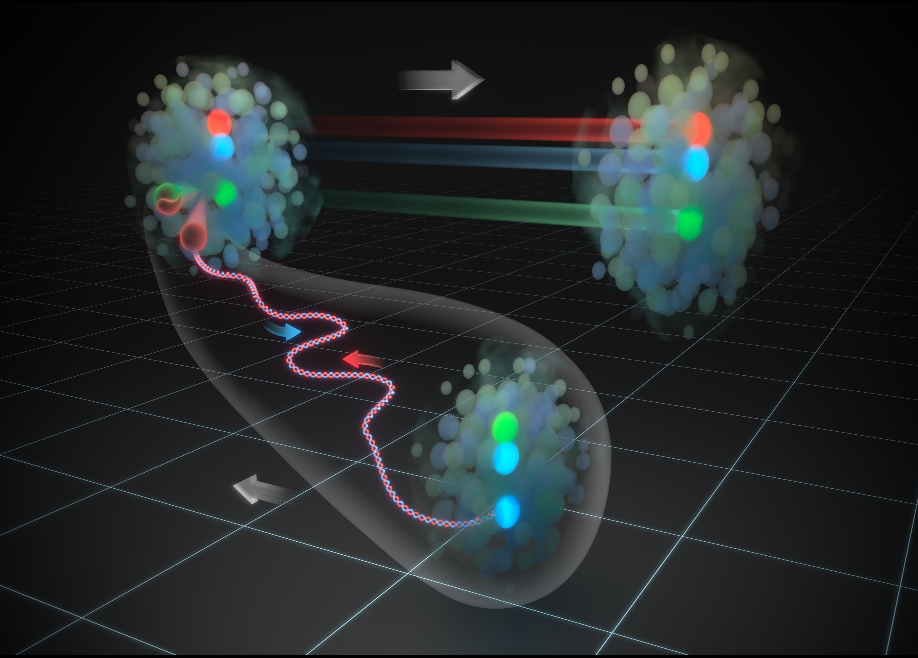}\\
\end{center}
  \caption{The new mechanism for emission of ``diffractive'' protons through color exchange which we proposed in Ref.~\cite{g}. The proton exchanges a color octet (gluon) with the quark-antiquark pair from the sea of the other proton. Consequently, the untouched valence color singlet follows its path as as fast proton.
The red and blue arrows illustrate the exchange of color. This visualization comes from Ref.~\cite{p} (source: IFJ PAN/DualColor).}
 \label{difpr}
\vspace{0.2cm}
\end{figure}

In our work~\cite{g} we proposed one well-defined realization of this idea\footnote{The Gluon Exchange Model (GEM), see Ref.~\cite{g} for more details.}, based on the assumption that the specific event type is defined by the number of color octets (gluons) exchanged between the projectile and nucleons from the target nucleus. Consequently, in the projectile hemisphere one can distinguish several classes of events, which differ very significantly in terms of momenta of final state baryons:

\begin{enumerate}
\item[{\bf 1.}]
{\em The final state proton is created from \underline{three} valence quarks from the projectile}; such secondary protons are very fast; their momentum distribution mirrors the sea quark momentum distribution. Such events are known as ``inelastic diffraction''. In Ref.~\cite{g} we demonstrated that the exchange of the gluon between the valence quarks from one and the quark-antiquark pair from the other proton provides a good description of the experimental proton ``diffractive peak'' from
Ref.~\cite{x}. This situation is presented in Fig.~\ref{difpr}.
\item[{\bf 2.}]
{\em The secondary nucleon (proton or neutron) originates from the fragmentation of the chain in which the diquark is made by two valence quarks from the projectile}; this event class corresponds to the situation originally postulated by the DPM~\cite{CapellaTranh,MJ1985}.
\item[{\bf 3.}]
{\em The secondary nucleon originates from the fragmentation of the chain in which the diquark is made from one valence quark and one quark from the sea}. Such a situation, which we will label as ``effective diquark'' following our considerations made in Sec.~\ref{sec35}, may occur in a multiple proton-nucleon collision. Evidently, the secondary nucleons will have lower typical momenta with respect to pt.~{\bf 2.} above.
\item[{\bf 4.}]
{\em There is no diquark in the projectile hemisphere}; in such a situation with large probability, there will be also no baryon in the projectile hemisphere. This situation we will label as ``decuplet exchange'', and a detailed justification will be presented in Sec.~\ref{sec35}.
\end{enumerate}

It is to be noted that the relative contributions of the different classes may vary
as a function of the atomic mass $A$, or the centrality of the $pA$ collisions. As a result, one can obtain a strong dependence of the final state nucleon momentum on the number of wounded nucleons, see Fig.~\ref{X2}. Our quantitative model description of $pp$ and $pC$ reactions from the latter figure can be found in Refs.~\cite{g} and~\cite{b}, respectively.
Here we only stress that our approach, including classes {\bf 1.}-{\bf 4.}~provides a reasonable description of the total distribution of net non-strange baryons in the projectile hemisphere of $pC$ collisions. Isospin effects (differences between proton and neutron spectra) appear surprisingly strong in the precise $pC$ data~\cite{y} and are {\em partially} explained by our model (see Ref.~\cite{b}). We will further address this subject in Sec.~\ref{sec4}.
Notwithstanding, several remarks should be made:
\begin{enumerate}
\item[$\bullet$]
The classes {\bf 1.}~and {\bf 3.}~above could not be realized without the presence of sea constituents. Specifically, the important class~{\bf 1.}~is realized by the presence of a sea quark and antiquark in a {\em single} proton-proton collision. This points at the importance of a properly complete Fock space available to the participating protons and nucleons, or more generally of the proper understanding of the partonic structure of the proton for a reliable description of the collision process.
\item[$\bullet$]
The classes~{\bf 3.}~and~{\bf 4.}~correspond to the disintegration of the (valence) diquark in multiple proton-nucleon collisions. From our three works~\cite{g,e,b} the latter appears as absolutely necessary in order to explain the experimental $pA$ data. Already for the ``multiple'' part of $pC$ reactions with $\langle n\rangle\approx 2.6$ wounded target nucleons, Fig.~\ref{fig1}, the lower limit for this process varies from one third to one half depending on the model applied. This in turn brings two questions:
\begin{enumerate}
\item[$\bullet$]
{\em The existence of the diquark}, that is, the issue whether the latter is to be understood as anything more than a $3^*$ color composition of two valence quarks, which therefore could be easily destroyed by the exchange of more than one color octet (gluon) as we explained in Refs.~\cite{g,e};
\item[$\bullet$]
{\em The preservation of the diquark} in the Wounded Constituent model~\cite{bialasbzdak1,bialasbzdak2,bialasbzdak3}, which apparently cannot be reconciled with the conclusions obtained in the framework of our approach.
\end{enumerate}
\end{enumerate}

%
%
%

\section{Color}
\label{sec35}

At this point it is necessary to discuss the role of color algebra which, in view of our works on Dual Parton and Gluon Exchange models~\cite{MJ1985,g,e,b}, we consider 
at the basic origin of baryon stopping phenomena in soft collisions. We underline that this discussion is in a great part a direct repetition of that made in Ref.~\cite{e} which must be also made here for consistency. The modifications with respect to the cited reference reflect the evolution of our understanding, resulting from a more in-depth analysis of experimental data~\cite{x,y} which we later performed in Refs.~\cite{g,b}. The present discussion can 
also 
be 
regarded as a late continuation of works~\cite{jjadach,jjapb}, which to the best of our knowledge were the first to explain the importance of different color configurations for valence quarks in high energy hadron-hadron reactions. We note that the cited works state the ``hidden'' character of color in hadrons in dual models, and argue that the inclusion of the junction line (Rossi and Veneziano, Ref.~\cite{rv}) in the quark model frame is in fact superfluous. Both issues are of importance for the considerations made below.
\subsection{Color of constituents in hadron-hadron collisions}
Both DPM and GEM follow the QCD description of ``hard'' processes where hadrons are described as systems of constituents (partons). The very important difference is that the number of QCD partons is infinite while that of constituents in DPM/GEM remains always finite. Thus {\em e.g.}, in a meson-baryon ($\pi p$) collision, the meson will be described as composed of a valence quark (color triplet) and valence antiquark (color antitriplet). The baryon will be a compound of a valence diquark $D$ and valence quark $q$. As we demonstrated in Ref.~\cite{g}, the inclusion of higher Fock states into the description (the hadron being composed not only of valence but also sea constituents) leads to a successful description of processes up to now known as ``inelastic diffraction''.

The incoming hadrons are color-neutral singlets before the collision. As the two valence constituents of the meson are a color triplet and antitriplet, only a color octet and color singlet can be made of them; thus the SU(3) color algebra  allows only for color singlet and color octet exchanges in meson-baryon reactions. In Ref.~\cite{e} we still claimed it plausible that the color singlet initiated diffractive processes, but this statement is at present far less certain as in Ref.~\cite{g} we found that the latter can be obtained from color octet exchange. Therefore we will concentrate on the latter below. After the exchange of one color octet between the particle~1~(meson) and~2~(baryon), both particles are in the color octet state. Color confinement enforces the formation of color singlets but these can form only by connecting constituents from {\em different} particles. These singlets will take the form of color field tubes (``strings'' or ``chains'') due to the fact that the included constituents move in opposite directions at relativistic velocities in the collision c.m.~system. 
\subsection{The diquark}
\label{secdq}
The baryon in the collision discussed above is made of three valence quarks. A very important item is how this fact is to be treated in the general case of a single 
baryon-baryon
collision. The product of three triplets will give four irreducible representations, namely the singlet, two octets, and the decuplet. However, the decuplet exchange is not allowed between these two hadrons which are initially both in color singlet state. 
%
%
Importantly, as a consequence, in 
such a
collision the baryon can be taken as a compound of only {\em two} valence constituents - the quark $(q)$ and the diquark $(D)$.

The above point is of very high importance for the entire understanding of the baryon transport process. From the point of view of SU(3) color algebra, the diquark appears as an inevitable consequence of the color configurations available in single 
baryon-baryon
collisions where the strong interaction proceeds through color (gluon) exchange. {\em No need appears at any point to consider the diquark as any kind of ``stable'' nor ``bound'' state}. Specifically, it may well be that the diquark ``exists'' only as a random configuration of two quarks forming a color antitriplet. At the moment when the possibility appears, 
it can disintegrate.
Such a possibility will be provided by color decuplet exchange not allowed in single 
baryon-baryon interactions,
but allowed in hadron-nucleus processes as 
it
will be 
discussed
below.

Prior to this, it is useful to compare meson-meson and baryon-baryon collisions. After the exchange of a gluon (color octet), both collisions will result in the formation of two color singlets of the color triplet-color antitriplet $(3-\bar{3})$ form. In both cases these singlets will take the shape of longitudinally extended ``chains'' as explained above. For meson-meson collisions each of them will be of the quark-antiquark $(q-\bar{q})$ type while for baryon-baryon collisions each of the two singlets will be of the quark-diquark type\footnote{In the above the lowest-lying Fock state of the baryon with no contribution from the sea is assumed.}. In our Ref.~\cite{e} we introduced the naming convention $D-q$ for the diquark belonging to the projectile and $q-D$ to the target particle. However, what said above, enforced by experimental data and our analyses~\cite{g,e,b}, forces us to somewhat redefine the physical understanding of this convention. Indeed, $D$ in both chains above does not anymore appear as a ``dynamical constituent'' of the proton as it was the case for the original Dual Parton Model, but only {\em as an ``effective shortcut'' for two quarks in a color antitriplet state} (valence $qq$, valence-sea $qq_s$, quarks from two different nucleons, etc.). This fact has numerous implications which we are only beginning to realize (some of them are exemplified and discussed in Refs.~\cite{g,b}). Here we come back to the Wounded Constituent Model where the latter diquark appears to us as a fully standalone object contributing to particle production in a way formally similar to the quark, and reiterate our intuition that the latter model may have problems in describing the strong baryon stopping in $pC$ collisions just as it was the case for our original (diquark-preserving) version of DPM as shown in Fig.~\ref{X2}. An independent analysis of this phenomenon by the authors of the latter model would greatly contribute to improving the understanding of the fate of participating quarks in the collision.
%
\begin{figure}[t]
\vspace{0.2cm}
\begin{center}
\hspace*{0.1cm}
\includegraphics[width=10.2cm]{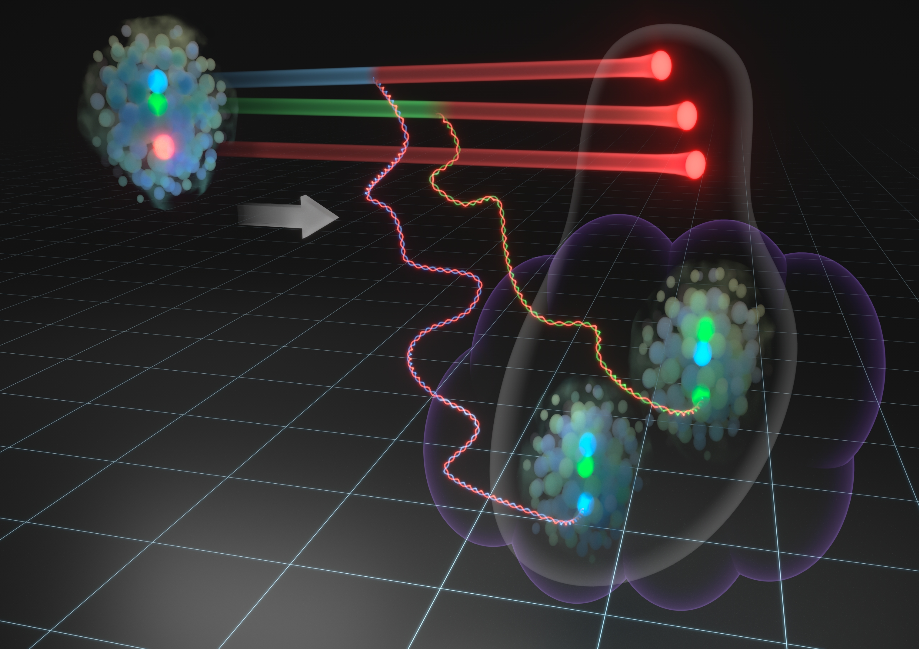}\\
\end{center}
  \caption{Mechanism for disintegration of the diquark which we introduced in Refs.~\cite{g,e}. After the color decuplet exchange in the collision of the proton with two nucleons in the carbon nucleus, the three valence quarks are found in a color-symmetric state which cannot anymore be described as quark+diquark system. This visualization comes from Ref.~\cite{p} (source: IFJ PAN/DualColor).}
 \label{dqdispr}
\vspace{0.2cm}
\end{figure}
%

\subsection{Color in collisions of protons with nuclei}
Let us now turn to the $pA$ collisions and the scattering of the proton on more than one nucleon. We note that for the somewhat unique case of Refs.~\cite{x,y} such collisions can be reliably extracted from experimental data as we did it in Refs.~\cite{g,e}, see Figs.~\ref{fig1} and~\ref{X2}. In such collisions a color decuplet exchange is not anymore forbidden by the rules of color algebra which creates a {\em qualitative difference with respect to single hadron-hadron interactions}. After this exchange the three constituent quarks in the proton are found in color decuplet state which is color-symmetric, and cannot be anymore described as a quark+diquark system. Thus, 
{\em there is no diquark in the projectile hemisphere.}
An artistic view of this situation is presented in Fig.~\ref{dqdispr}. 
%

This situation gives an excellent example for a more general finding:
for two or more gluons exchanged between the constituents of the projectile and wounded nucleons in the nucleus, {\em new diagrams} appear 
which were not available in the original DPM model
(see {\em e.g.} 
our Ref.~\cite{g}).
The color decuplet exchange from Fig.~\ref{dqdispr} 
will build
the event class~{\bf 4.} 
mentioned
in Sec.~\ref{sec3}. It 
will lead 
to the formation of chains of the $q-D$ type (where $D$ is either ``the original valence diquark in a wounded target nucleon'' or ``a diquark made by two quarks from two nucleons'', both descriptions having anyway only an effective meaning as explained in Sec.~\ref{secdq}). This issue is discussed at length in Ref.~\cite{b}. Modulo some exotic and essentially unphysical scenarios addressed therein, the most logical expectation is that such chains will induce very long transfers of baryon number towards the nucleus (diquark) hemisphere. Indeed, our analysis~\cite{b} suggests the presence of this diagram as one of the sources of the strong baryon stopping observed in Fig.~\ref{X2}. 
It is to be underlined, however, that another important contribution arises from an ``effective diquark'' composed of one valence and one sea quark in the projectile. This latter contribution corresponds to event class {\bf 3.} in Sec.~\ref{sec3}. As it is described in Ref.~\cite{b}, the corresponding transfers of baryon number in rapidity are longer than the original DPM mechanism, but shorter than these induced by color decuplet exchange.

These new contributions to baryon stopping described above are 
in our view at the origin of the failure of the original DPM (and likely any similar diquark-preserving model), as it was shown in Fig.~\ref{X2}. They are in fact also at the origin of the old controversy between the nuclear stopping power extracted by Busza and Goldhaber from experimental data~\cite{Busza} and model simulations made by Je\.zabek and R\'o\.za\'nska~\cite{comments}. Unfortunately, at the time the experimental data were insufficient to provide a reliable understanding of the underlying baryon stopping process, a situation which is now at last improved with the new experimental data~\cite{x,y} which 
served as 
basis for our phenomenological analyses~\cite{g,e,b}. We believe that in this way the puzzle of nuclear stopping power has been solved.




\subsection{Collisions of antiprotons with protons and nuclei}

Let us now address antiproton-induced collisions. For $\bar{p}p$ collisions the color octet exchange will result in the formation of two chains, respectively of $\bar{q}-q$ and $\overline{D}-D$ type. 
The latter chain will contribute mainly to multiparticle final states including one fast antibaryon and one fast baryon in opposite hemispheres. Another possible channel exists with mesons only in the final state, which contributes to antiproton annihilation at high energies.
%
For the $\bar{p}p$ case, color decuplet exchange is not forbidden by the rules of color algebra. Thus, an exciting possibility is that this process would not be very much suppressed with collision energy, leading to another contribution to baryon number annihilation. 

Finally, let us consider the case most interesting in our opinion, that is the exchange of the color decuplet in the collision of the antiproton with multiple nucleons. Following our findings discussed above, it seems difficult not to expect the contribution from this diagram to $\bar{p}A$ reactions at high energy. The resulting formation of chains of the $q-\bar{q}$ type can be meant as {\em annihilation of the antiproton negative baryon number over many different nucleons in the nucleus}, in principle very efficient in removing the projectile baryon number from the projectile hemisphere. The latter contribution should be dependent on the size of the colliding nucleus. As the occurrence of such a process would be very highly informative on the fate of partons in the collision {\em and}  on the correctness of our approach from Refs.~\cite{g,e,b}, this subject justifies in our mind new measurements of antiproton-induced collisions as it will be discussed in Sec.~\ref{sec4}.


\section{Chain fragmentation}
\label{sec36}

Presently we will address the second item of importance which, in terms of time evolution of the collision comes {\em after} what was described in the precedent section, namely the fragmentation of the chain into hadrons and most particularly baryons in the final state. As said above, a chain is a non-perturbative color singlet created by the exchange of color in the collision, which takes the form of a longitudinally extended color field tube. In our studies described here we adapted a fully phenomenological approach and implemented the mechanism of chain fragmentation in the form of fragmentation and ``isospin flip'' functions directly on the basis of data from proton-proton collisions. We note that this approach, being in fact a direct adaptation of the present ``standard'' proposed in Ref.~\cite{MJ1985} to more precise experimental data, permitted an excellent description of diffractive and non-diffractive proton and neutron spectra in $pp$ reactions~\cite{g}. The technical aspects of this approach are described at length in Ref.~\cite{b}.

\begin{figure}[t]
\vspace{0.2cm}
\begin{center}
\includegraphics[width=6.2cm]{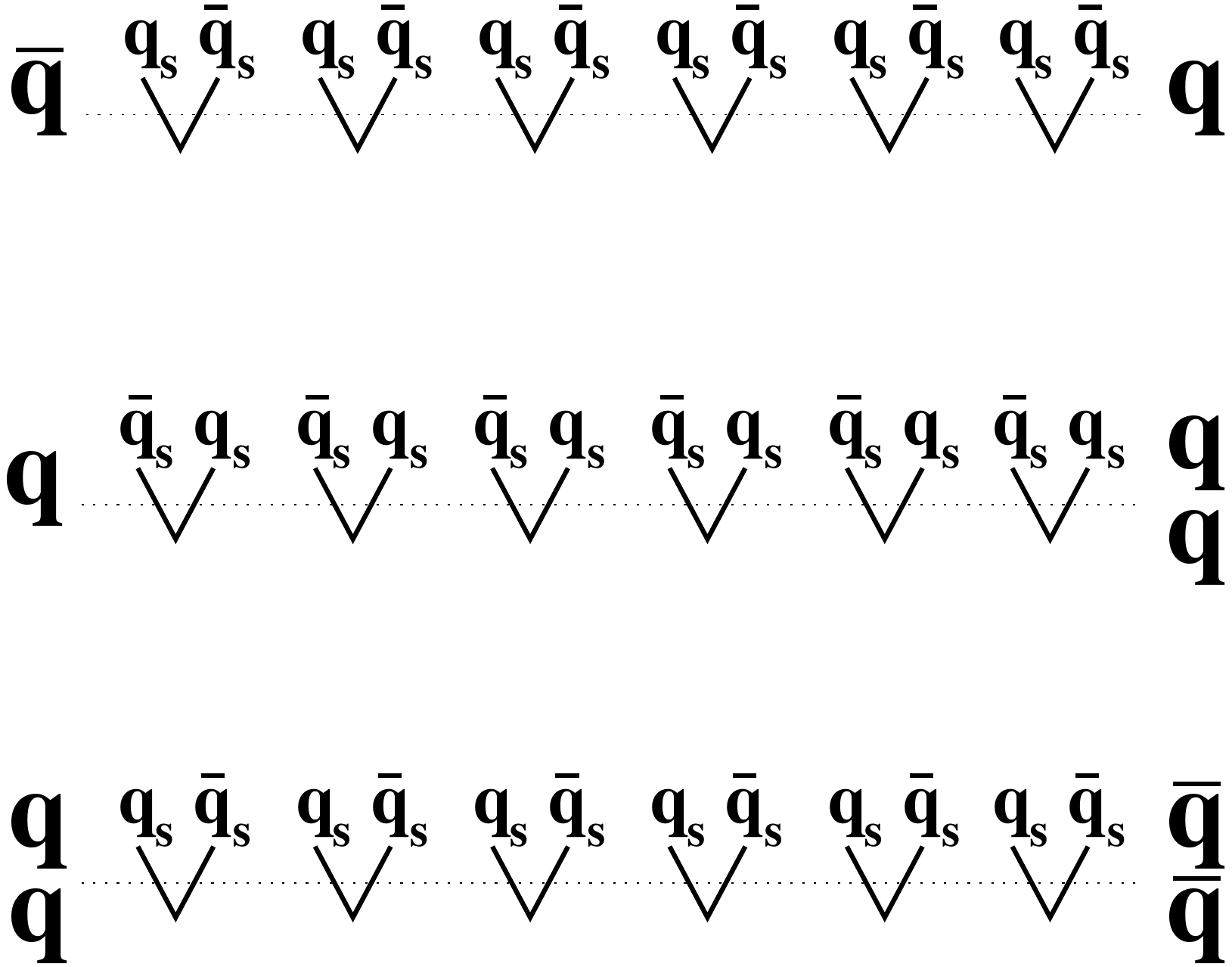}\\
\end{center}
\vspace*{0.2cm}
  \caption{Schematic drawing of chain fragmentation by creation of quark-antiquark pairs, shown for the case of the $q-\bar{q}$, $D-q$, and $\overline{D}-D$ chains.}
 \label{Y}
\vspace{0.2cm}
\end{figure}

 Nevertheless, it is necessary to explain our picture of the physical chain fragmentation process which is independent on given technical implementation, and in particular the assumptions which we believe need to be made in order to obtain a reliable description of experimental data. Here the main assumption is that {\em the color string which connects the ends of the chain breaks predominantly by creation of a quark-antiquark pair} as it is drawn in Fig.~\ref{Y}. This assumption implies that creation of a diquark-antidiquark pair (independently of the meaning of the ``diquark'' concept as discussed in Sec.~\ref{sec35}) is strongly suppressed. We believe that this is justified by the observation that antibaryons are produced dominantly in the central region of nucleon-nucleon collisions, see {\em e.g.} Ref.~\cite{x}, whereas contributions to string fragmentation due to diquark-antidiquark pair creation would be much broader in rapidity. An additional,
even stronger justification are the low values of antibaryon/pion ratios seen in the cited experimental data. 

The second assumption is the inherent {\em isospin-symmetry} of the chain fragmentation process, which implies equal probabilities for production of $u\bar{u}$ and $d\bar{d}$ pairs.
This assumption implies that the isospin effects in particle production are in fact induced by the {\em initial conditions} of the string fragmentation process, that is the initial isospin (flavor) of the constituents of the incoming protons (nucleons), see Fig.~\ref{Y}. For the time being we conclude that the success of our approach in describing the very strong isospin effects in forward proton and neutron emission in $pp$ collisions~\cite{g} confirms the correctness of this assumption.\\

It is useful to consider three basic cases of chain fragmentation\footnote{This part of the discussion is again largely a repetition of that made in Ref.~\cite{e}.}:
\begin{enumerate}[leftmargin=*]
\item[(1)]
in {\em meson-meson collisions}, the basic ``building block'' in the collision will be the chain of $q-\bar{q}$ type, characterized by a given effective mass. This will fragment into hadrons in the same way as a $W$ boson, or virtual photon of the same mass. Following our nomenclature mentioned in Sec.~\ref{sec35}, in the meson-meson collision two chains are formed, $q-\bar{q}$ and $\bar{q}-q$. With elapsing time both these chains will break by creation of quark-antiquark pairs, and fragment dominantly into mesons. This specific mechanism of chain fragmentation implies that the electric charge of an initial constituent will be transferred into a particle whose rapidity will not be much different from the rapidity of this constituent. This means that during the chain fragmentation process, the electric charge is deposited {\em locally}, within a limited distance in rapidity. The electric charge of leading (most energetic) particles will depend on the electric charge of the corresponding constituent. For the transport of the initial baryon number 
along 
this chain of $3-\bar{3}$ type, the baryon number 1/3 \mbox{(-1/3)} will be transferred from the triplet (antitriplet) end, 
in the direction of the other end of the string. 
%
\item[(2)]
in {\em proton-proton collisions}, the basic chain is of $D-q$ type, notwithstanding the fact that $q-\bar{q}_s$ chains can form in the emission of ``diffractive'' protons through color exchange, see Ref.~\cite{g}. For $D-q$ chains no analogy can be drawn to $W$ bosons or virtual photons as above and the characteristics of the fragmentation {\em must} be extracted from experimental data. Our studies~\cite{b} show that the quantitative reliability of this extraction remains doubtful if the latter do not include simultaneously protons and neutrons which, to the best of our knowledge, was the case only for our recent analysis described in the cited reference. 
The basic features of fragmentation of the $D-q$ chain, confirmed by the this analysis, can be described as follows. The general picture of propagation 
of the initial baryon number along the chain can 
be seen at first
as the transport of the initial 1/3 from the quark end of the string towards the initial 2/3 at the diquark end of the string, see Fig.~\ref{Y}. More precisely, 
with the chain fragmenting predominantly through production of $q \bar{q}$ pairs as explained above, the largest probability density will correspond to fragmentation into a baryon in vicinity of
%
%
the original accumulation (2/3) resulting from the presence of two quarks at the end of the string. The transport of net baryon number towards more central rapidities will occur with 
rapidly
decreasing probability.
Importantly, all of the latter will be valid also for configurations in which there is no valence diquark left but only an effective object made of two essentially random quarks from different nucleons or from the sea, like the classes~{\bf 3.}~and~{\bf 4.}~in Sec.~\ref{sec3} above. Specifically, in {\em proton-nucleus collisions}, the exchange of the color decuplet and resulting disappearance of any possible ``diquark'' color configuration will assign the incoming valence quarks to three strings of $q-D$ type. Following what was said above, each of these will transport the initial baryon number 1/3 in the direction of the 2/3 from the target nucleon, providing thus an extremely strong mechanism for baryon stopping.

One more important remark should still be added on the fragmentation of the $D-q$ chain. As it emerges from our conclusion of the ``unstable'' or even purely effective character of the diquark strongly supported by experimental data (Fig.~\ref{X2}), the ``microscopic'' picture of transport of baryon number in the 
course
of string fragmentation by creation of $q\bar{q}$ pairs can be seen as composed of two processes. First, the (fast) final state baryon can be created from the {\em original} valence diquark in vicinity of its rapidity (Fig.~\ref{Y}). Secondly, however, it may well be that the (effective or weakly bound) diquark will be disintegrated in the course of the fragmentation process, {\em i.e.} that one of its valence quarks will be substituted by a newly created quark from the sea. In this latter case the transport of baryon number towards central rapidities will consist in the propagation of the diquark through consecutive substitutions, along the string\footnote{This is reminiscent of the discussion of baryon transport in multiperipheral-type models made by Jadach in Ref.~\cite{jadach-mult}.}.

\item[(3)]
in {\em antiproton-proton collisions}, the two basic chains which form are of $\bar{q}-q$ type, fragmenting in the way described in (1) above, and the new triplet-antitriplet $(\overline{D}-D)$ chain (Fig.~\ref{Y}). As it has already been considered in Ref.~\cite{MJ1985}, at large collision energies the fragmentation of the latter should be similar to that of the $D-q$ chain from 2., and more exactly it should appear as the transport of the initial baryon number -2/3 and 2/3 from both ends towards the center of the string. At lower collision energies the string will become shorter in terms of rapidity, and annihilation of baryon number will occur. The energy-dependence of this annihilation contribution is quite evident as the typical length of the chains in rapidity is directly connected to $\sqrt{s}$. Color decuplet exchange, which following our findings 
would be already active in $pA$ collisions, is not permitted in $pp$ but is allowed in $\bar{p}p$ reactions. The resulting formation of three $\bar{q}-q$ chains, fragmenting mostly into mesons as explained above, would of course create another contribution to baryon number annihilation with a possibly different energy-dependence than that of the $\overline{D}-D$ chain. The presence of that contribution in $\bar{p}p$ reactions would need to be elucidated. On the other hand, as we already said above, the apparent frequent disintegration of the diquark in multiple proton-nucleon collisions which we need to explain the experimental data~\cite{g} makes it difficult to believe that this process will be absent in $\bar{p}A$ reactions, which may lead again to assigning all the valence antiquarks in the $\bar{p}$ to different strings of  $\bar{q}-q$ type. Similarly to (2) above, this mechanism we consider as a very powerful tool for removing the projectile (negative) baryon number from the projectile hemisphere, linked this time with baryon number annihilation.
\end{enumerate}
It is evident that the schemes described above can be realized in somewhat different technical implementations, not uniquely limited to our - rather successful - adaptation of the scheme~\cite{MJ1985} as described in Ref.~\cite{b}. 
Of course, a better quantitative understanding of the process of string fragmentation will result in more restrictive descriptions and will increase the predictive power of the resulting model. Work in this direction is in progress and will be reported elsewhere.



\section{Implications and future studies}
\label{sec4}

Summarizing the present status of our phenomenological work on baryon data~\cite{x,y} brings the following picture:
\begin{enumerate}
\item[$\bullet$]
Modern experimental data with simultaneous proton and neutron measurement in the entire projectile hemisphere play an absolutely essential role in constraining the postulated phenomenology, revising the earlier wrong assumptions, and drawing firm conclusions on the process of baryon transport in high energy collisions.
\item[$\bullet$]
Both spectra of ``diffractive'' and ``non-diffractive'' baryons can be explained by the process of color octet exchange.
\item[$\bullet$]
Baryon spectra in $pA$ reactions can be explained only when a very significant amount of events where the valence diquark is disintegrated is assumed; this statement is valid already for the proton colliding with $n=2$ nucleons. Consequently, the diquark would be merely a $3^*$ color configuration of two quarks, easily disintegrated by the exchange of two or more gluons.
\item[$\bullet$]
The inclusion of diagrams including the latter disintegration (classes {\bf 3.}~and {\bf 4.}~from Sec.~\ref{sec3}) allows for a reasonable description of the total net (non-strange) baryon spectrum in the projectile hemisphere of multiple proton-nucleon collisions ($pC$ reactions where the proton collides with two or more nucleons,
see
Fig.~\ref{X2}).
\item[$\bullet$]
The subdivision of non-strange baryons into net proton ($p-\overline{p}$) and net neutron ($n-\overline{n}$) spectra shows that $pp$ collisions are characterized by strong isospin effects especially at forward rapidity. Importantly {\em the same statement is valid for multiple proton-nucleon collisions}, where isospin effects remain of comparable magnitude. This is evidently informative on the fate of quarks (u and d) in the collision, and contradicts the naive sequential model which we addressed in Sec.~\ref{sec1} above. Our approach presented in Sec.~\ref{sec3} also predicts sizeable isospin effects in multiple proton-nucleon collisions, but even 
these are somewhat underestimated with respect to what is obtained from experimental data~\cite{x,y}; this points in our view at the need of a further 
improvement of our model~\cite{g,b}.
\end{enumerate}
The present situation appears to us as already significantly more clear than that emerging from earlier studies~\cite{Busza,comments,MJZalewski,Brenner,Bailey,MJ1985,buszaledoux} as far as the role of quarks and gluons, color exchange, the concept of the diquark, and chain fragmentation into baryons including isospin effects are concerned. Several issues, including very reasonable but wrong assumptions left from these studies could be clarified and corrected. The ``instability'' or even purely effective character of the diquark, Sec.~\ref{secdq}, creates an interesting controversy with the Wounded Constituent Model, very successful in the description of charged particle spectra in different types of collisions~\cite{bialasbzdak1,bialasbzdak2,bialasbzdak3}. Further improvement of the understanding of the role of constituent partons in different aspects of non-perturbative high energy reactions is to be expected once this controversy is solved. 

It is quite evident that the starting point for this progress is the modern experimental data~\cite{x,y}, far more restrictive in constraining and falsifying model assumptions thanks to excellent coverage, well controlled systematics, and most of all the simultaneous measurement of protons {\em and} neutrons which allows to get hold of the total baryon number and of isospin (quark flavor) effects\footnote{Modulo a small contribution from strange baryon which has been estimated elsewhere~\cite{Varga2003}.}. At first sight it may seem surprising that experimental data of such completeness and quality remain in fact very scarce. With the exception of preliminary $pPb$ data from NA49~\cite{hgfqm,epi2002}, we are actually not aware of experimental measurements for protons and neutrons simultaneously, characterized by similarly large coverage in phase space. This is notwithstanding the evident fact that very numerous datasets on protons exist from the very numerous experiments performed up to now (a partial but very useful bibliography including detailed comparative studies of NA49 with older datasets can be found Refs.~\cite{x,y}, including also some neutron measurements). A more in-depth consideration brings easily the explanation for this situation. On one hand, this lies in the extremely large and very high quality expert effort necessary to obtain such datasets, requiring in practice many years (more than a decade) of expert work. In spite of the very high scientific value, such an effort is difficult to reconcile with the practical limitations imposed to scientists in terms of stable funding, career development, pressure for rapid publication, etc. On the other hand, the second reason lies in the limitations imposed to the strongest high energy scientific community that is LHC experiments, in terms of unavoidable difficulties in continuous coverage of the entire phase space up to beam rapidity for collision energies of several TeV in the c.m.s.

In this context, it is useful to examine the possibilities for improvement of this situation, that is of obtaining new experimental data on baryon stopping in hadron-hadron and hadron-nucleus reaction in the mid-term future, of quality comparable to these cited in the present paper. We see two main directions for such studies:
\begin{enumerate}[leftmargin=*]
\item[(i)] {\em hadron-hadron and hadron-nucleus collisions for different nuclei}, including in particular baryonic and mesonic beams, as well as heavier nuclei like $Pb$ - this direction can be considered as permitting a direct extension of our studies made in Refs.~\cite{g,e,b}, and
\item[(ii)] {\em antiproton-proton and antiproton-nucleus measurements} which, as discussed at length in this paper, would permit the use of baryon number annihilation for the realization of the new diagrams with disintegrated diquark, highly informative on the collision process\footnote{We note that this idea was proposed already in Ref.~\cite{MJZalewski}, see Ref.~\cite{Bailey} for comparison.}.
\end{enumerate}
At the present moment two main options are apparent to us.
 
\begin{enumerate}[leftmargin=*]
\item[-]
{\em NA61/SHINE}. The NA61/SHINE experiment~\cite{ji} is a logical option for new measurements of the type of~\cite{x,y} as it can be considered, on the technical level, largely as a continuation of NA49~\cite{nim}. The experiment already obtained final data on proton distributions in $pp$, $pC$, $\pi^+Be$, $\pi^+C$, and $BeBe$ collisions~\cite{na61-2,na61-1,na61-3,na61-4}. The detector is 
evidently characterized by a large coverage in rapidity in the projectile hemisphere of the reaction. A first proposal for future measurements with antiprotonic beams has already been 
made 
by us~\cite{ar}. On the technical level, it is at present under consideration by the NA61/SHINE Collaboration
and has been reported to the CERN SPS Committee in Ref.~\cite{s}. An important issue to be clarified is whether the NA61/SHINE Projectile Spectator Detector (PSD~\cite{ji}) can be used for neutron measurements in a way similar to the NA49 hadronic calorimeter~\cite{x,y}. Up to now the PSD has been used for centrality estimation via spectator energy deposit which, similarly to the NA49 device, in principle does not preclude further applications for hadron-induced collisions. Following an optimistic scenario, it is therefore not excluded that future datasets from NA61/SHINE could provide a significant contribution to a further improvement of understanding of the baryon stopping process.
\item[-] {\em AFTER@LHC.} 
As we already said above, the LHC energy range seems sub-optimal to us in view of new measurements that could have ``generic'' impact on the understanding of the baryon stopping process, mostly due to unavoidable difficulties in ensuring a homogeneous coverage of the available phase space at collision c.m.~energies of the order of many TeV. What is more, such high energies bring the additional, important complication of very abundant baryon-antibaryon pair production, which results in the 
determination
of net baryon 
emission
becoming insecure.
Summing up, the LHC energy range imposes a situation far more challenging that the SPS energy range.
On the other hand, measurements performed in the framework of the LHC community benefit from non-negligible practical advantages in terms of size of available  manpower, detector technology, extended multi-step quality assessment procedures, and most of all relatively very high statistics, especially for ``bulk'' features of 
the collision like
the
$dn/dy$ spectra discussed in the present paper. Consequently any option of even auxiliary baryon stopping measurements is to be examined carefully. Of particular potential interest is the AFTER@LHC program~\cite{al} where already at the present moment, interesting and encouraging fixed-target results on antiproton distributions at $\sqrt{s}=110$~GeV are available from LHCb~\cite{a}. Of course the fixed target configuration reduces, by decreasing the collision c.m.s.~energy, also the problem of overwhelming pair production discussed above, and at the same time increases the (relative) experimental coverage of the total available phase-space. The comparison of coverage in rapidity for the different experiments presented in Ref.~\cite{al} suggests the possibility of valuable, high statistics measurements also in the baryon stopping domain. While in principle, these cannot be compared with the completeness of proton/neutron measurements~\cite{x,y} discussed in the present paper, they may prove essential once it comes to verification of the approach proposed in~\cite{g,e,b} in terms of collision
energy dependence. An interesting issue for further clarification is to what extent would this be possible to perform measurements with, {\em e.g.}, light isospin-symmetric nuclei which bring no difference between proton and neutron emission.

\section{Summary and conclusions}

Modern, high coverage experimental data on proton and neutron emission in proton-proton and proton-nucleus reactions allow for a new ``res\-tart'' of baryon stopping studies in the sense of these undertaken many decades ago, and lifting up some of the important limitations of the earlier analyses. With our proposal of the GEM (Gluon Exchange Model) backed up by such modern experimental data, it appeared possible to clarify or better elucidate several issues left from earlier baryon stopping studies. In particular, this included providing a unified, homogeneous description of ``diffractive'' and ``non-diffractive'' baryon emission by considering a properly complete ensemble of Fock states, a better understanding of strong isospin effects in proton/neutron emission and consequently falsification of early phenomenological assumptions on the latter, and fixing the old controversy on the nuclear stopping power provided by the DPM approach with respect to data-based analyses by Busza and Goldhaber. The latter issue is connected to our finding of very frequent diquark disintegration already in the collision of the proton with two nucleons in the nucleus, which in turn sheds a new light on the possibly purely effective character of the concept of the ``diquark''. This creates an interesting controversy with respect to the Wounded Constituent Model which once clarified, may lead to a further improvement in understanding of the connection between baryon stopping and the process of non-perturbative charged particle production. Furthermore, these studies bring interesting implications for new baryon number annihilation diagrams in antiproton-induced collisions. Consequently, an evident need becomes apparent for new high quality experimental data on proton, neutron and strange baryon emission in collisions of protons, antiprotons and mesons with hadrons and atomic nuclei.


\end{enumerate}

\begin{center}
{\bf Acknowledgments}\\
\end{center}
%


M.J. gratefully acknowledges the many years long collaboration with Maria R\'o\.za\'nska which contributed in an essential way to the phenomenology and ideas presented in this paper.
Also A.R. warmly thanks M.~R\'o\.za\'nska for very instructive discussions
on the physics of soft baryon emission.

We thank Andrzej Bia\l{}as for many interesting and useful comments to this paper.


This work was supported by the National Science Centre, Poland
(grant no.~2014/14/E/ST2/00018).


\begin{thebibliography}{99}

\bibitem{jj}
T.~Jaroszewicz and M.~Je\.zabek,
{\it Z. Phys. } \textbf{C4}, 277 (1980).


\bibitem{m-blan}
L.~Landau, I.~Pomeranchuk, translations RT-530 and RT-864. Translated from {\em Dokl. Akad. Nauk SSR} {\bf 92}, 535 (1953) and {\bf 92}, 735 (1953).

\bibitem{m-b2}
M.~Miesowicz, {\it Acta Phys. Pol.} \textbf{B33}, 105 (1972).  

\bibitem{m-b1}
A.~Bia\l{}as,
{\it Acta Phys. Polon.} \textbf{B15}, 647 (1984).

\bibitem{bialas-stodolsky-apb}
A.~Bia\l{}as and L.~Stodolsky,
{\it Acta Phys. Polon.} \textbf{B7}, 845 (1976).

\bibitem{stodolsky}
L.~Stodolsky,
{\it Phys. Rev. Lett.} \textbf{28}, 60 (1972).


\bibitem{x}
T.~Anticic \textit{et al.} [NA49 Collaboration],
{\it Eur. Phys. J.} {\bf C65}, 9 (2010).

\bibitem{y}
B.~Baatar \textit{et al.} [NA49 Collaboration],
{\it Eur. Phys. J.} {\bf C73}, 
2364 (2013).

\bibitem{g}
M.~Je\.zabek and A.~Rybicki,
{\it Phys. Lett. } \textbf{B816}, 136200 (2021).

\bibitem{pcdiscus}
G.~Barr \textit{et al.},
{\it Eur. Phys. J.} {\bf C49}, 919 (2007).

\bibitem{bialasczyz}
A.~Bia\l{}as, W.~Czy\.z and W.~Furma\'n{}ski,
{\it Acta Phys. Polon.} \textbf{B8}, 585 (1977).


\bibitem{Busza}
W.~Busza and A.~S.~Goldhaber,
{\it Phys. Lett.} {\bf B139}, 235 (1984).

\bibitem{bialasbzdak1}
A.~Bia\l{}as and A.~Bzdak,
{\it Phys. Lett.} {\bf B649}, 263 (2007).

\bibitem{bialasbzdak2}
A.~Bia\l{}as and A.~Bzdak,
{\it Phys. Rev.} {\bf C77}, 034908 (2008).

\bibitem{bialasbzdak3}
M.~Barej, A.~Bzdak and P.~Gutowski,
{\it Phys. Rev.} {\bf C100}, 
064902 (2019).

\bibitem{CapellaTranh}
A.~Capella and J.~Tran Thanh Van,
{\it Phys. Lett.} {\bf B93}, 146 (1980).

\bibitem{comments}
M.~Je\.zabek and M.~R\'o\.za\'nska,
{\it Phys. Lett.}  \textbf{B175}, 206
(1986).

\bibitem{bgk}
S.~J.~Brodsky, J.~F.~Gunion and J.~H.~K\"uhn,
{\it Phys. Rev. Lett.} \textbf{39}, 1120 (1977).


\bibitem{e}
M.~Je\.zabek and A.~Rybicki,
{\it Acta Phys. Polon. } \textbf{B51}, 
1207
(2020).

\bibitem{b}
M.~Je\.zabek and A.~Rybicki,
[arXiv:2105.13741 [nucl-th]].

\bibitem{MJZalewski}
M.~Je\.zabek and K.~Zalewski,
{\it Acta Phys. Polon.} {\bf B11}, 425 (1980).


\bibitem{Brenner}
A.~Brenner \textit{et al.},
{\it Phys. Rev.} {\bf D26}, 1497 (1982).

\bibitem{Bailey}
R.~Bailey \textit{et al.},
{\it Z. Phys.} {\bf C29}, 1 (1985).

\bibitem{MJ1985}
M.~Je\.zabek, J.~Karczmarczuk and M.~R\'o\.za\'nska,
{\it Z. Phys.} {\bf C29}, 55 (1985).


\bibitem{buszaledoux}
W.~Busza and R.~Ledoux,
{\it Ann. Rev. Nucl. Part. Sci.} {\bf 38}, 119 (1988).

\bibitem{p}
Press release, IFJ PAN Press Office, 9 June 2021.\\
 {\em https://press.ifj.edu.pl/en/news/2021/06/09/}



\bibitem{jjadach}
S.~Jadach and M.~Je\.zabek,
{\it Phys. Lett. }\textbf{B80}, 295 (1979).

\bibitem{jjapb}
S.~Jadach and M.~Je\.zabek,
{\it Acta Phys. Polon.}  \textbf{B10}, 715 (1979).

\bibitem{rv}
G.~C.~Rossi and G.~Veneziano,
{\it Nucl. Phys.} \textbf{B123}, 507 (1977).

\bibitem{jadach-mult}
S.~Jadach,
{\it Nucl. Phys.} \textbf{B99}, 514 (1975).

\bibitem{hgfqm}
H.~G.~Fischer [NA49 Collaboration],
{\it Nucl. Phys.} {\bf A715}, 118 (2003).

\bibitem{epi2002}
A.~Rybicki [NA49 Collaboration],
{\it Acta Phys. Polon.} {\bf B33}, 1483 (2002).

\bibitem{ji}
N.~Abgrall \textit{et al.} [NA61/SHINE Collaboration],
{\it JINST} \textbf{9}, P06005 (2014).

\bibitem{Varga2003}
D.~Varga [NA49 Collaboration],
{\it Eur. Phys. J.} {\bf C33}, S515 (2004).

\bibitem{nim}
S.~Afanasiev \textit{et al.} [NA49 Collaboration],
{\it Nucl. Instrum. Meth.} {\bf A430}, 210 (1999).


\bibitem{na61-2}
A.~Aduszkiewicz \textit{et al.} [NA61/SHINE Collaboration],
{\it Eur. Phys. J.} {\bf C77}, 
671 (2017).

\bibitem{na61-1}
N.~Abgrall \textit{et al.} [NA61/SHINE Collaboration],
{\it Eur. Phys. J.} {\bf C76}, 
84 (2016).

\bibitem{na61-3}
A.~Aduszkiewicz \textit{et al.} [NA61/SHINE  Collaboration],
{\it Phys. Rev.} {\bf D100}, 
112004 (2019).


\bibitem{na61-4}
A.~Acharya \textit{et al.} [NA61/SHINE Collaboration],
{\it Eur. Phys. J.} \textbf{C81}, 
73 (2021).

\bibitem{ar}
A.~Rybicki, talk at the NA61/SHINE Collaboration Meeting, 23 September 2020.

\bibitem{s}
A. Acharya \textit{et al.}, [NA61/SHINE Collaboration], {\it CERN-SPSC-2020-023}.










\bibitem{al}
C.~Hadjidakis, D.~Kiko\l{}a, J.~P.~Lansberg, L.~Massacrier, M.~G.~Echevarria, A.~Kusina, I.~Schienbein, J.~Seixas, H.~S.~Shao,
A.~Signori, \textit{et al.},
{\it Phys. Rept.} \textbf{911}, 1 (2021).


\bibitem{a}
R.~Aaij \textit{et al.} [LHCb Collaboration],
{\it Phys. Rev. Lett.} \textbf{121}, 
222001 (2018).




















\end{thebibliography}
\end{document}